\documentclass[useAMS,usenatbib,fleqn]{mnras}
\usepackage{multirow}
\usepackage{graphicx}
\usepackage{adjustbox}
\usepackage{color,soul}
\usepackage{epsfig}
\usepackage{amssymb,amsmath}
\usepackage{mathtools}
\usepackage{aas_macros}
\usepackage{fixltx2e}
\title[Wandering SMBH Census]{
Origins and Demographics of Wandering Black Holes
}

\author[Ricarte et al.]{Angelo Ricarte$^{1,2}$, Michael Tremmel$^{3}$, Priyamvada Natarajan$^{3,4}$, 
\newauthor
Charlotte Zimmer$^{5}$, Thomas Quinn$^{6}$
\\
$^{1}$ Center for Astrophysics | Harvard \& Smithsonian, 60 Garden Street, Cambridge, MA 02138, USA \\
$^{2}$ Black Hole Initiative at Harvard University, 20 Garden Street, Cambridge, MA 02138, USA \\
$^{3}$ Department of Astronomy, Yale University, 52 Hillhouse Avenue, New Haven, CT 06511, USA \\
$^{4}$ Department of Physics, Yale University, P.O. Box 208121, New Haven, CT 06520, USA \\
$^{5}$ Department of Economics, Yale University, P.O. Box 208268, New Haven, CT 06520, USA \\
$^{6}$ Department of Astronomy, University of Washington, PO Box 351580, Seattle, WA 98195, USA
}

\date{\today}

\begin{document}
\pagerange{\pageref{firstpage}--\pageref{lastpage}} \pubyear{2021}
\maketitle

\begin{abstract}
We characterise the population of wandering black holes, defined as those physically offset from their halo centres, in the {\sc Romulus} cosmological simulations. Unlike most other currently available cosmological simulations, black holes are seeded based on local gas properties and are permitted to evolve dynamically without being fixed at halo centres.  Tracking these black holes allows us to make robust predictions about the offset population.  We find that the number of wandering black holes scales roughly linearly with the halo mass, such that we expect thousands of wandering black holes in galaxy cluster halos.  Locally, these wanderers account for around 10 per cent of the local black hole mass budget once seed masses are accounted for.  Yet for higher redshifts ($z\gtrsim 4$), wandering black holes both outweigh and outshine their central supermassive counterparts.  Most wandering black holes, we find, remain close to the seed mass and originate from the centres of previously disrupted satellite galaxies.  While most do not retain a resolved stellar counterpart, those that do are situated farther out at larger fractions of the virial radius.  Wanderers with higher luminosities are preferentially at lower radius, more massive, and either closer to their host's mid-planes or associated with a stellar overdensity.  This analysis shows that our current census of supermassive black holes is incomplete and that a substantial population of off-centre wanderers likely exists.
\end{abstract}


\section{Introduction}

Every massive galaxy is believed to host a supermassive black hole (SMBH) at its centre \citep{Kormendy&Richstone1995,Kormendy&Ho2013}.  SMBH masses correlate with properties of the inner regions of their host galaxies \citep{Ferrarese&Merritt2000,Gebhardt+2000,Tremaine+2002}, which can be understood as the result of SMBH-galaxy co-evolution over cosmic time \citep[e.g.,][]{Haehnelt+1998,Kauffmann&Haehnelt2000,Volonteri+2003,Somerville+2008,Natarajan2014}.  In the contemporary galaxy formation paradigm, SMBHs grow primarily from gas that makes its way to the galactic centre, resulting in active galactic nuclei (AGN).  Subsequently, feedback from the growing AGN helps quench star formation via radiation, winds, and/or jets \citep{DiMatteo+2005,Springel+2005,Croton+2006,Sijacki+2007}.  Modern cosmological simulations now self-consistently evolve star formation and SMBH growth from the early universe to the present day using sub-grid models for unresolved astrophysical processes, such as SMBH fuelling and feedback \citep{Schaye+2015,Dubois+2016,Bocquet+2016,Tremmel+2017,Pillepich+2018,Dave+2019}.  

Hierarchical structure formation, whereby larger halos result from the mergers of smaller ones, naturally produces SMBHs that are offset from the centres of their galaxies. The journey from galactic scales to the eventual SMBH merger in the inner regions of galaxies requires several different processes across a large dynamic range of spatial scales \citep{Begelman+1980,Colpi2014}.  Only those secondary SMBHs which complete their journeys to the centres of their galaxies will form binaries with the primary SMBH, which can continue to evolve into the merger regime.  It is possible that in some cases the orbital decay of these SMBHs can become stalled on scales much too large to form a bound binary \citep{Governato+1994,Schneider+2002,Volonteri&Perna2005,Callegari+2009,Callegari+2011,Bellovary+2010,Micic+2011,Dosopoulou+Antonini2017,Dvorkin+Barausse2017,Tamfal+2018,Tremmel+2018b,Bellovary+2021}.  

First, on kilo-parsec scales, dynamical friction is the primary mechanism for angular momentum loss \citep{Chandrasekhar1943}.  For low enough SMBH masses, the dynamical friction timescale can exceed the Hubble time.  In addition, there has been a growing awareness that additional delays of Gyrs or longer can occur at this step, especially in clumpy high-redshift galaxies as noted in simulations \citep{Biernacki+2017,Tremmel+2018a,Pfister+2019,Bortolas+2020,Ma+2021}.  Accreting SMBHs wandering on kilo-parsec scales can potentially be observed as dual/offset AGN \citep[e.g.,][]{Comerford+2009,Comerford+2013,Mezcua+2020,Reines+2020} or offset hyperluminous X-ray sources (HLXs) \citep{King&Dehnen2005,Barrows+2019}. There is also a growing sample of SMBH candidates observed in galactic halos within ultra-compact dwarfs, which may represent the stripped remnants of once-massive galaxies \citep{Seth+14,Ahn+2017,Afanasiev+2018,Voggel+2019}

In spherical symmetry, dynamical friction from stars becomes less efficient on parsec scales resulting in what has often been referred to as the ``final parsec problem'' \citep{Milosavljevic&Merritt2001,Milosavljevic&Merritt2003}.  However, a wide variety of mechanisms has been proposed to overcome this barrier, including gas dynamical friction \citep{Ostriker1999,Escala+2004,Armitage&Natarajan2005,Mayer+2007}, individual gas clumps \citep{Fiacconi+2013,Goicovic+2017}, spherical asymmetry \citep{Khan+2013,Rantala+2017,Gualandris+2017}. and even multi-body interactions with other SMBHs \citep{Ryu+2018}.  

These journeys culminate in gravitational wave emission, which can be detected by pulsar timing arrays and the planned Laser Interferometer Space Antenna (LISA) mission \citep{Hobbs+2010,LISA+2017,NANOGrav2020}.  Delays between halo merger and SMBH merger as discussed above can significantly impact gravitational wave event rates \citep{Barausse+2020}.  While there are several other sources of uncertainties in the current determination of LISA merger event rates from SMBH mergers, including those arising from accretion physics, and the hitherto unknown formation mechanisms for initial black hole seeds \citep{Sesana+2007,Klein+2016,Ricarte&Natarajan2018b,Dayal+2019}, the role of SMBH dynamics prior to reaching the gravitational wave regime is one of the key unknowns.  The SMBH merger rate can also have a significant impact on SMBH mass and spin evolution, especially in the most massive galaxies \citep{Volonteri+2005,Volonteri+2013}.  Several models predict that SMBH mergers can contribute significantly to the mass budgets of the most massive low-redshift SMBHs, those hosted by gas-poor ellipticals \citep{Kulier+2015,Ricarte&Natarajan2018a,Pacucci&Loeb2020}. 

Cosmological simulations can be powerful tools for predicting the formation of SMBH binaries because they naturally include realistic merger histories and self-consistently evolving galaxies and large-scale structure.  The {\sc Romulus} cosmological simulations \citep{Tremmel+2017} are one of the only large-scale simulations that self-consistently tracks the orbital evolution of SMBH pairs down to sub-kpc scales through the use of a sub-grid dynamical friction prescription to account for missing dynamical friction from stars and dark matter (but not gas) \citep{Tremmel+2015}.  {\sc Romulus} is therefore uniquely able to predict which SMBHs, following a galaxy merger, will actually make it to the centre of their new host galaxy and how long that process will take. Indeed, as presented in \citet{Tremmel+2018a}, {\sc Romulus} predicts that many SMBH binaries form after several Gyrs of orbital evolution, while some SMBHs will never make it to the centre. As a result, Milky Way-mass galaxies in {\sc Romulus} are found to host an average of 12 SMBHs, which typically wander the halo far from galactic centre \citep{Tremmel+2018b}.

In this work, we perform a detailed study of the {\sc Romulus} suite of simulations to characterise the population of wandering SMBHs beyond the centres of their host halos.  The outline of paper is as follows: in \S\ref{sec:methodology}, we provide a brief summary of the {\sc Romulus} simulations, focusing on the SMBH physics and the post-processing necessary to perform this study. In \S\ref{sec:results}, we present the results of this analysis and present the detailed properties of this population:  bulk properties in \S\ref{sec:bulk}, intra-halo statistics in \S\ref{sec:intra-halo}, and a study of stellar counterparts in \S\ref{sec:stellar_counterparts}.  We discuss the caveats and broader implications of this study in \S\ref{sec:discussion}, and summarise our results in \S\ref{sec:conclusions}.

\section{Methodology}
\label{sec:methodology}

In this work, we present results from analysis of the {\sc Romulus} suite of simulations, both {\sc Romulus25} and {\sc RomulusC}.

\subsection{The {\sc Romulus} Simulations}
\label{sec:romulus}

Here, we summarise the important aspects of the SMBH physics implemented in the {\sc Romulus} simulations that is relevant to our current discussion, further details can be found in \citet{Tremmel+2017,Tremmel+2019}.  The {\sc Romulus} simulations consist of {\sc Romulus25}, a (25 Mpc)$^3$ volume, and {\sc RomulusC}, a zoom-in on a $10^{14} \ \mathrm{M}_\odot$ galaxy cluster.  These simulations utilise the Tree + Smoothed Particle Hydrodynamics (SPH) code {\sc ChaNGa} \citep{Menon+2015,Wadsley+2017}, with a dark matter particle mass of $3.39 \times 10^5 \ \mathrm{M}_\odot$ and a gas particle mass of $2.12 \times 10^5 \ \mathrm{M}_\odot$.  

SMBHs in both simulations are seeded based on local gas properties, without any knowledge of the host halo or galaxy.  As a consequence, not all halos are explicitly guaranteed to host a SMBH, and multiple SMBHs may form in the same halo.  A gas particle that would have formed a star instead is transformed into a SMBH if all of the following hold true:
\begin{itemize}
    \item It is very metal poor, with metallicity $Z<3\times10^{-4}$.
    \item It is very dense, at least 15 times the star forming threshold of 0.2 $m_p/\mathrm{cc}$.
    \item It is warm, with a temperature between 9500 and 10000 K.
\end{itemize}
This seeding prescription emulates the direct collapse black hole (DCBH) scenario, where the high temperatures and low metallicities allow for a large Jeans mass, which can directly collapse into a BH seed \citep{Oh&Haiman2002,Bromm&Loeb2003,Lodato&Natarajan2006,Begelman+2006}.  SMBHs are seeded with an initial mass of $10^6 \ \mathrm{M}_\odot$ to ensure that they are always more massive than other particles in the simulation, which could otherwise cause spurious scattering events.  Due to this local seeding prescription, SMBHs will occasionally form very close to each other and promptly merge, but this does not occur for the majority of BHs. As discussed in \citet{Tremmel+2018a}, the majority of wandering SMBHs in Milky-Way mass halos have not experienced any mergers, nor have grown beyond twice their seed mass.  The vast majority ($>90$ per cent) of SMBHs are seeded at redshifts $z\geq 5$, well prior to epochs studied in this work. At the lower redshifts studied here, the inferred properties of the wandering population are no longer dominated by the initial conditions pertinent to their formation. 

Gravity in the Romulus suite is softened below 350 pc using a spline kernel, equivalent to 250 pc Plummer softening length.  The dynamical friction force onto SMBHs is corrected for the underestimate due to this softening, and this is done by integrating the \citet{Chandrasekhar1943} dynamical friction equation within the smoothing radius \citep{Tremmel+2015}.  With the addition of this force correction, SMBHs are allowed to orbit freely without being explicitly pinned to the centres of their halos. Note that the {\sc Romulus} simulations do not account for dynamical friction from gas \citep{Ostriker1999}. Such gas drag has been included in other large-scale simulations, but without the effects of stars and dark matter \citep{Dubois+2014}. The effect of gas may become particluarly important at high redshift \citep[though see recent results from][for further discussion]{Pfister+2019}. 

Two SMBHs merge if they are within two gravitational softening lengths (within 0.7 kpc) and have low enough relative velocities to be considered bound.  That is, if $\frac{1}{2}\Delta \vec{v} < \Delta \vec{a} \cdot \Delta \vec{r}$, where $\Delta \vec{v}$ is the relative velocity, $\Delta \vec{a}$ is the relative acceleration, and $\Delta \vec{r}$ is the relative displacement of two SMBHs \citep{Bellovary+2011}.

Further, in these simulations, SMBHs accrete based on the Bondi-Hoyle-Lyttleton formula \citep{Bondi1952}, modified to account for angular momentum support.  The accretion rate is given by 

\begin{equation}
    \dot{M}_\bullet = \alpha(n) 
    \begin{dcases*}
    \frac{\upi(GM_\bullet)^2\rho}{(v^2_\mathrm{bulk} + c^2_s)^{3/2}} & \text{if $v_\mathrm{bulk} > v_\theta$} \\ 
    \frac{\upi(GM_\bullet)^2\rho c_s}{(v^2_\theta + c^2_s)^{2}} & \text{if $v_\mathrm{bulk} < v_\theta$}, \\ 
    \end{dcases*}
\end{equation}

\noindent where $G$ is the gravitational constant, $M_\bullet$ is the SMBH mass, $\rho$ is the ambient mass density, $c_s$ is the ambient sound speed, $v_\theta$ is the local rotational velocity of surrounding gas, and $v_\mathrm{bulk}$ is the bulk velocity relative to the SMBH \citep[see][for additional details]{Tremmel+2017}.  Ambient quantities are calculated using the 32 closest gas particles.  The coefficient $\alpha(n)$ is a number density-dependent boost factor given by 

\begin{equation}
    \alpha(n) = 
    \begin{dcases*}
    \left( \frac{n}{n_{\mathrm{th},*}} \right)^2 & \text{if $n \geq n_{\mathrm{th},*}$} \\ 
    1 & \text{if $n < n_{\mathrm{th},*}$} \\ 
    \end{dcases*}
\end{equation}

\noindent where $n_{\mathrm{th},*}$ is the star formation number density threshold, meant to represent the limit beyond which the simulation no longer resolves the multi-phase interstellar medium \citep{Booth&Schaye2009}.

To model AGN feedback, thermal energy is isotropically imparted into nearby gas particles during accretion, assuming a radiative efficiency of 10 per cent, and a feedback coupling efficiency of 2 per cent.  There is no separate kinetic/radio mode included in the AGN feedback implementation.  Details of the sub-grid physical processes that describe star formation and feedback from the stellar population as well as the growing SMBH are described in \cite{Tremmel+2017,Tremmel+2019}.

\subsection{Post-processing: finding the wandering population}
\label{sec:post-processing}

We use the {\sc Amiga} halo finder to define halos and derive their virial masses, virial radii, and centres \citep{Knollmann&Knebe2009}.  We require each halo to have at least $10^4$ dark matter particles for it to be considered well-resolved. Outputs are processed using The Agile Numerical Galaxy Organisation System ({\sc TANGOS}) \citep{Pontzen&Tremmel2018}.  {\sc Amiga} assigns SMBHs to a halo, along with all other dark matter, star, and gas particles, based on whether or not the objects are bound to the structure. SMBHs may also be assigned to substructure within more massive halos. Halo centres are determined using the shrinking-spheres approach \citep{Power+2003}, which consistently traces the centres of central galaxies as well. The accuracy of the method is of order the gravitational softening length, within which gravity is not resolved and the potential minimum cannot be accurately defined. For {\sc Romulus} this is 350 pc. The calculation may be further complicated and uncertain during on-going halo/galaxy mergers of similar mass, although we do not expect this to systematically affect our results.

We define a SMBH to be ``central'' if it is within 0.7 kpc of the halo centre (twice the gravitational softening length), and ``wandering'' if it lies anywhere beyond this radius. This definition allows for multiple central SMBHs in the same halo, and also permits the most central, massive, or luminous SMBH in a halo to be designated as wandering. The minimum threshold of 0.7 kpc takes into account uncertainties in the SMBH dynamics, as well as the definition of the centre of the galaxy, due to the gravitational softening length. This is also the distance threshold within which two SMBHs are allowed to merge (due to the unreliable dynamics at these scales). Outside of 0.7 kpc we can be confident that gravity and the dynamical evolution is fully resolved and the SMBH particles are not about to merge within the simulation. To further study the wandering population, we consider three different radial cuts for their locations ranging from the most generous to a more stringent definition as follows:
\begin{itemize}
    \item All SMBHs beyond 0.7 kpc, such that the wandering SMBH population is the complement of the central SMBH population.
    \item All SMBHs beyond 5 kpc, a more conservative cut.
    \item All SMBHs beyond 0.1 $R_{200}$, as calculated for each halo.
\end{itemize}
The first cut is the most inclusive classification criteria, the second extends out to a physical scale that corresponds roughly the size scale of the visible galaxy, and the final cut is the tightest aligned with the most stringent definition for a wanderer.  

When we present data binned by halo mass, we combine halos from the {\sc Romulus25} (the field) and {\sc RomulusC} (the cluster) simulations by only using {\sc RomulusC} data to extend into halo mass bins for which {\sc Romulus25} has no data.  We refrain from including {\sc RomulusC} galaxies in less massive bins because environmental effects (such as tidal stripping and ram pressure stripping) significantly modify lower mass galaxies in the cluster environment compared to their counterparts in the field.  At $z=0.05$, the lowest redshift for which we analyse these simulations, the most massive halo in {\sc Romulus25} is a galaxy group with a mass of $2.1 \times 10^{13} \ \mathrm{M}_\odot$, while the main halo of cluster simulation {\sc RomulusC} has a mass of $1.4 \times 10^{14} \ \mathrm{M}_\odot$.

\section{Results}
\label{sec:results}

In \S\ref{sec:bulk}, we study bulk statistics of wandering SMBHs compared to their central counterparts, including their total masses and luminosities.  Then, in \S\ref{sec:intra-halo}, we then study the intra-halo statistics of the population, such as their mass function and radial distribution.  Lastly, in \S\ref{sec:stellar_counterparts}, we examine whether or not wandering SMBHs may reveal their presence via a stellar overdensity.

\subsection{Population Statistics}
\label{sec:bulk}

\subsubsection{Central and Wanderer Abundances}
\label{sec:numbers}

\begin{figure}
  \centering
  \adjincludegraphics[trim={0.2\width} {0.1\width} {0.2\width} {0.07\width},clip,width=0.45\textwidth]{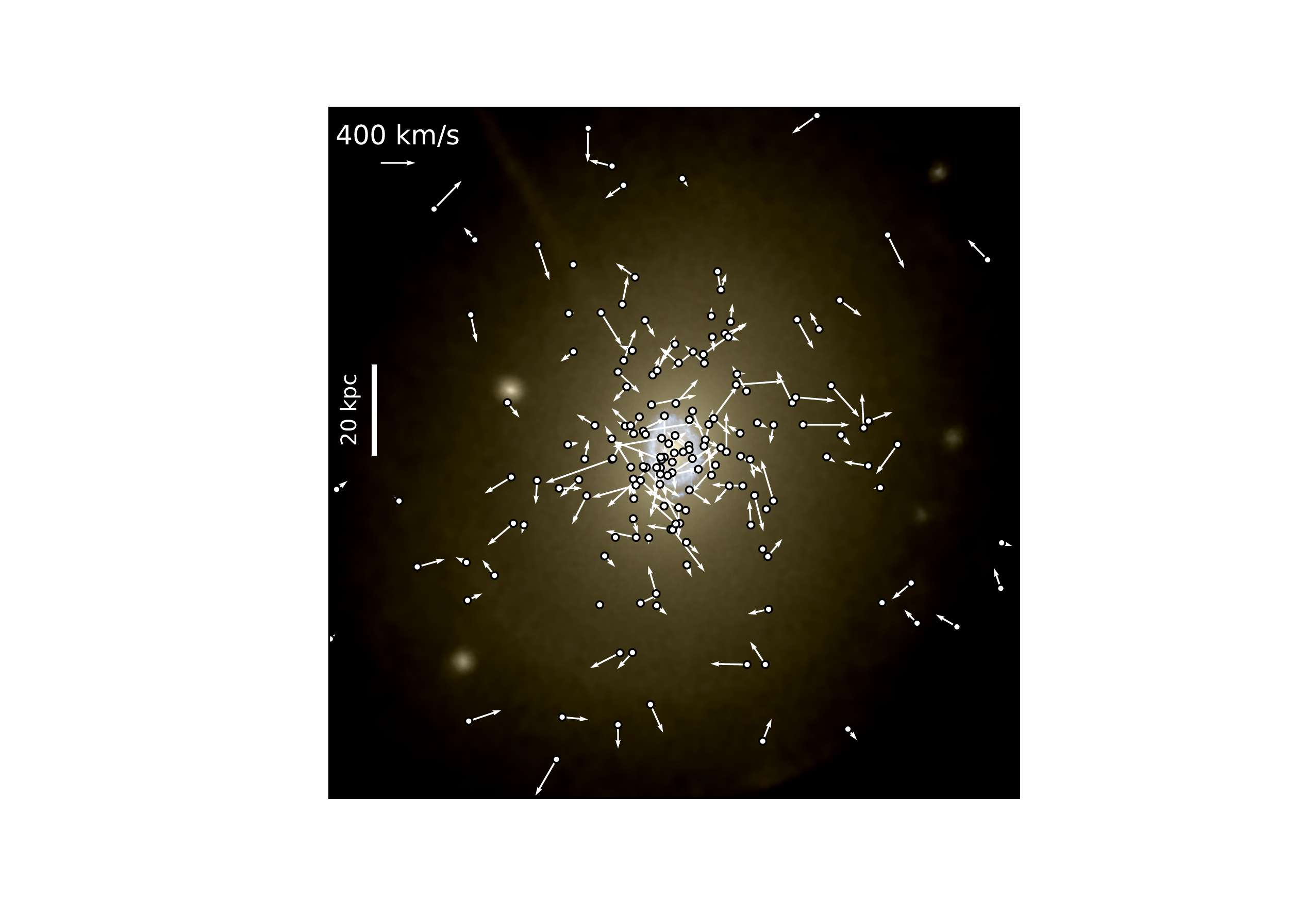} \\
  \adjincludegraphics[trim={0.2\width} {0.1\width} {0.2\width} {0.07\width},clip,width=0.45\textwidth]{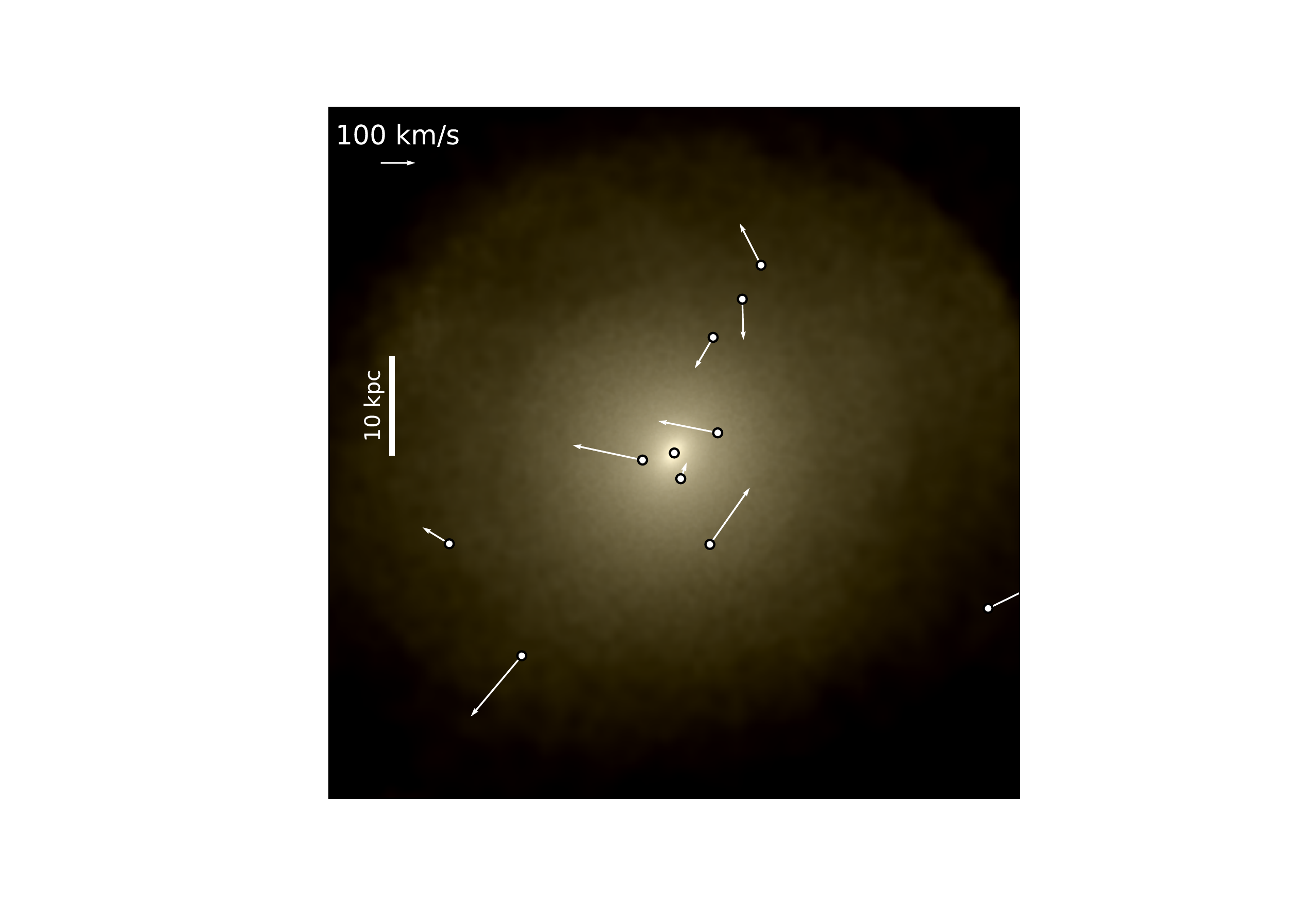}\\
  \adjincludegraphics[trim={0.2\width} {0.1\width} {0.2\width} {0.07\width},clip,width=0.45\textwidth]{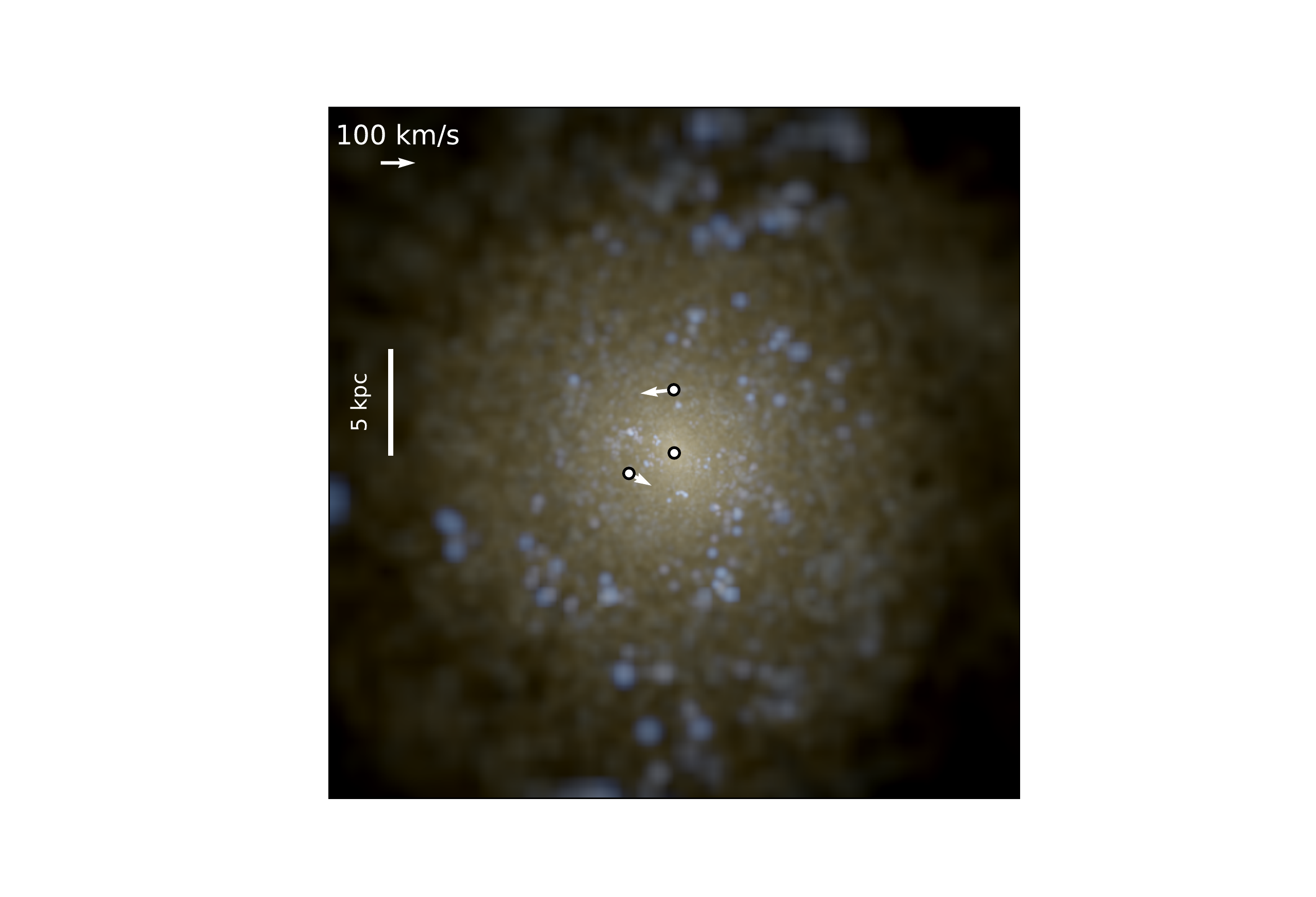}
  \caption{Examples of wandering SMBH populations of galaxies in {\sc Romulus25}, stepping through two decades in halo mass.  From top to bottom, these halos have virial masses of $2\times10^{13} \ \mathrm{M}_\odot$, $2\times10^{12} \ \mathrm{M}_\odot$, and $2\times10^{11} \ \mathrm{M}_\odot$ respectively.  The total number of wanderers in each galaxy is 241, 15 and 2 respectively, following a linear scaling between number of wanderers and halo mass.  Images are generated assuming a faint surface brightness limit of 28 $\mathrm{mag} \; \mathrm{arcsec}^{-2}$.  Note the different size and velocity scales in each panel. \label{fig:images}}
\end{figure}

Wandering SMBHs are ubiquitous in the {\sc Romulus} simulations, and occur in greater numbers in more massive halos.  In Figure \ref{fig:images}, we plot three sample images of galaxies selected to span two decades in halo mass.  From top to bottom, these galaxies reside in halos with virial masses of $2\times10^{13} \ \mathrm{M}_\odot$\footnote{The blue stars visible 
in this halo reflect a recent merger event.}, $2\times10^{12} \ \mathrm{M}_\odot$, and $2\times10^{11} \ \mathrm{M}_\odot$ respectively.  The location and velocity of each SMBH bound to the halo is marked with a circle and an associated arrow.  These halos host 241, 15, and 2 wanderers respectively (not all of which fit in the images shown), consistent with a linear trend with halo mass.  As we further explore in \S\ref{sec:radial_distribution}, wandering SMBHs span a broad range of distances, extending all the way out to the virial radius.  These images are generated assuming a surface brightness limit of 28 $\mathrm{mag} \; \mathrm{arcsec}^{-2}$ to better visualise the stellar halo.  Even at this faint surface brightness limit, the wandering SMBHs are not associated with stellar counterparts.

\begin{figure*}
  \centering
  \includegraphics[width=\textwidth]{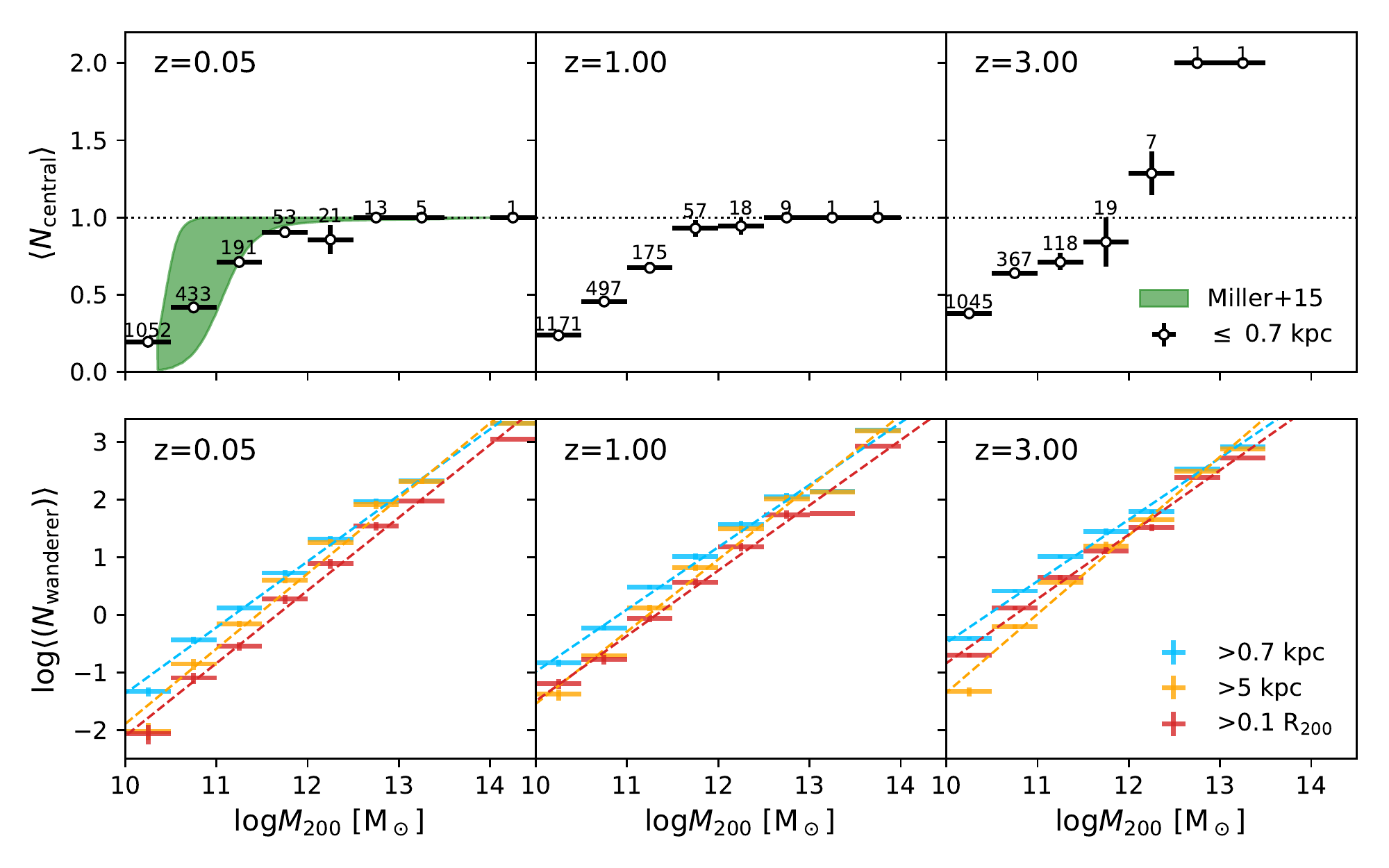}
  \caption{Average numbers of central SMBHs (above) and wandering SMBHs (below) as a function of halo mass for three different redshifts in {\sc Romulus}.  Different colours reflect different cuts of the wandering population, while the numbers above the error bars in the top row are the number of unique halos in each bin.  We compare the number of central SMBHs at the lowest redshift with the occupation fraction estimate of \citet{Miller+2015}, after applying the stellar-to-halo mass relation of \citet{Moster+2013} and find remarkable agreement despite the seeding mechanism lacking any knowledge of global halo properties.  Interestingly, we find that the most massive halos at $z=3$ sometimes host an extra central SMBH, requiring that they have too large relative velocities to be considered bound.  A power law fit to the number of wandering SMBHs as a function of halo mass is plotted using dashed lines.  The number of wanderers scales roughly linearly with the halo mass at all redshifts.  At all epochs, wandering SMBHs far outnumber central SMBHs in all except dwarf galaxy halos. \label{fig:occupation}}
\end{figure*}

In Figure \ref{fig:occupation}, we plot as a function of halo mass the average number of central SMBHs in the top row, and the average number of wandering SMBHs (using the three different definitions outlined in \S\ref{sec:post-processing}) in the bottom row for three different redshifts.  We write the number of halos that exist in each bin above of the error bars in the top row.  \citet{Miller+2015} estimate the SMBH occupation fraction at $z=0$ as a function of stellar mass, and we overplot their observational results in green after applying the stellar-to-halo mass relation of \citet{Moster+2013}.  In this figure, and subsequent figures in this section, error bars in the vertical direction enclose the 16th to 84th percentile regions regions (meant to represent 1$\sigma$), estimated via bootstrapping the halos in a given mass bin.

There is a remarkable level of agreement between the average number of centrals in {\sc Romulus} and the locally observed occupation fraction.  Recall that the SMBH seeding prescription used in {\sc Romulus} is based on local gas properties, without any knowledge of global halo properties.  Almost all seeds are established by $z \gtrsim 5$, and the occupation fraction at $z=0$ depends on a combination of seeding, growth, and dynamical processes that bring SMBHs into galaxy centres.  It is thus notable that observations and predictions from {\sc Romulus} agree at this level despite the modelling uncertainties in all three of the these processes, which bolsters our confidence that {\sc Romulus} produces a reasonable population of wandering SMBHs.  The distributions at $z=1$ and $z=3$ follow the same general shape as the local bin, though interestingly, the average number of central SMBHs exceeds unity in the highest mass bins at $z=3$.  Recall that SMBHs merge in {\sc Romulus} not only if they are within two gravitational smoothing lengths, but also if they have small enough relative velocities to be considered bound.  We were surprised to find a few massive galaxies at $z=3$ in {\sc Romulus} that host multiple central SMBHs that obviously meet the spatial separation criterion but not the velocity criterion for merging, resulting in multiple centrals.

In the bottom row, we find that the number of wandering SMBHs exceeds unity for $M_{200} \gtrsim 10^{11} \ \mathrm{M}_\odot$ at all three epochs and that for larger halo masses the abundance of wanderers is significantly larger.  The number of wandering SMBHs scales roughly linearly with the halo mass, a trend stretching from dwarf galaxy halos to galaxy clusters.  We find additional structure if stellar mass bins are adopted instead, due to the non-linear mapping between stellar and halo mass, suggesting that halo mass is the more fundamental property.  A power-law regression (linear in logarithmic space) to the number of wanderers as a function of halo mass is plotted as the dashed line.  This linearity holds regardless of the distance cut from the galaxy centre chosen to define wanderers, although a steeper slope is obtained using the 5 kpc cut.  This is unsurprising as 5 kpc is an evolving fraction of the virial radius depending on the host halo mass and redshift.  At $z=0.05$, for the $>0.7$ kpc, $> 5$ kpc, and $>0.1 \ R_\mathrm{200}$ populations, we find:
$\log_{10}N_w = 1.15[\log_{10} (M_{200}/10^{12} \; \mathrm{M}_\odot)] + 0.93$, $\log_{10}N_w = 1.31[\log_{10} (M_{200}/10^{12} \; \mathrm{M}_\odot)] + 0.72$, and $\log_{10}N_w = 1.26[\log_{10} (M_{200}/10^{12} \; \mathrm{M}_\odot)] + 0.42$ respectively.  In low mass halos, the number of wanderers increases with increasing redshift, reflecting the redshift evolution of the SMBH occupation fraction.

In earlier work with these simulations, \citet{Tremmel+2018b} determined that Milky Way mass halos host $12.2 \pm 8.4$ SMBHs within their virial radii, consistent with these results.  Meanwhile, the main cluster mass halo of the cluster simulation {\sc RomulusC}, with a virial mass of $10^{14} \ \mathrm{M}_\odot$ at $z=0.05$, hosts a staggering 1613 wandering SMBHs, excluding those that are associated with individual galaxy scale subhalos.  We cautiously note here that halo finders have difficulty identifying substructure in dense galaxy clusters, such that some of these wandering SMBHs could potentially reside in undetected subhalos.  In an investigation of mass segregation, \citet{Joshi+2016} compared the {\sc Rockstar} and {\sc Amiga} halo finders and determined that the mass attributed to sub-halos was inconsistent between the two halo finders inside approximately 0.2 times the virial radius.  In the main halo of {\sc RomulusC}, 70 per cent of wandering SMBHs are within $0.2R_\mathrm{200}$ at $z=0.05$.  Nevertheless, the number of wanderers in the cluster halo is consistent with extrapolating the linear trend found at lower masses (as shown in the lower panel of Figure \ref{fig:occupation}).

\subsubsection{Central and Wanderer Masses}

\begin{figure*}
  \centering
  \includegraphics[width=\textwidth]{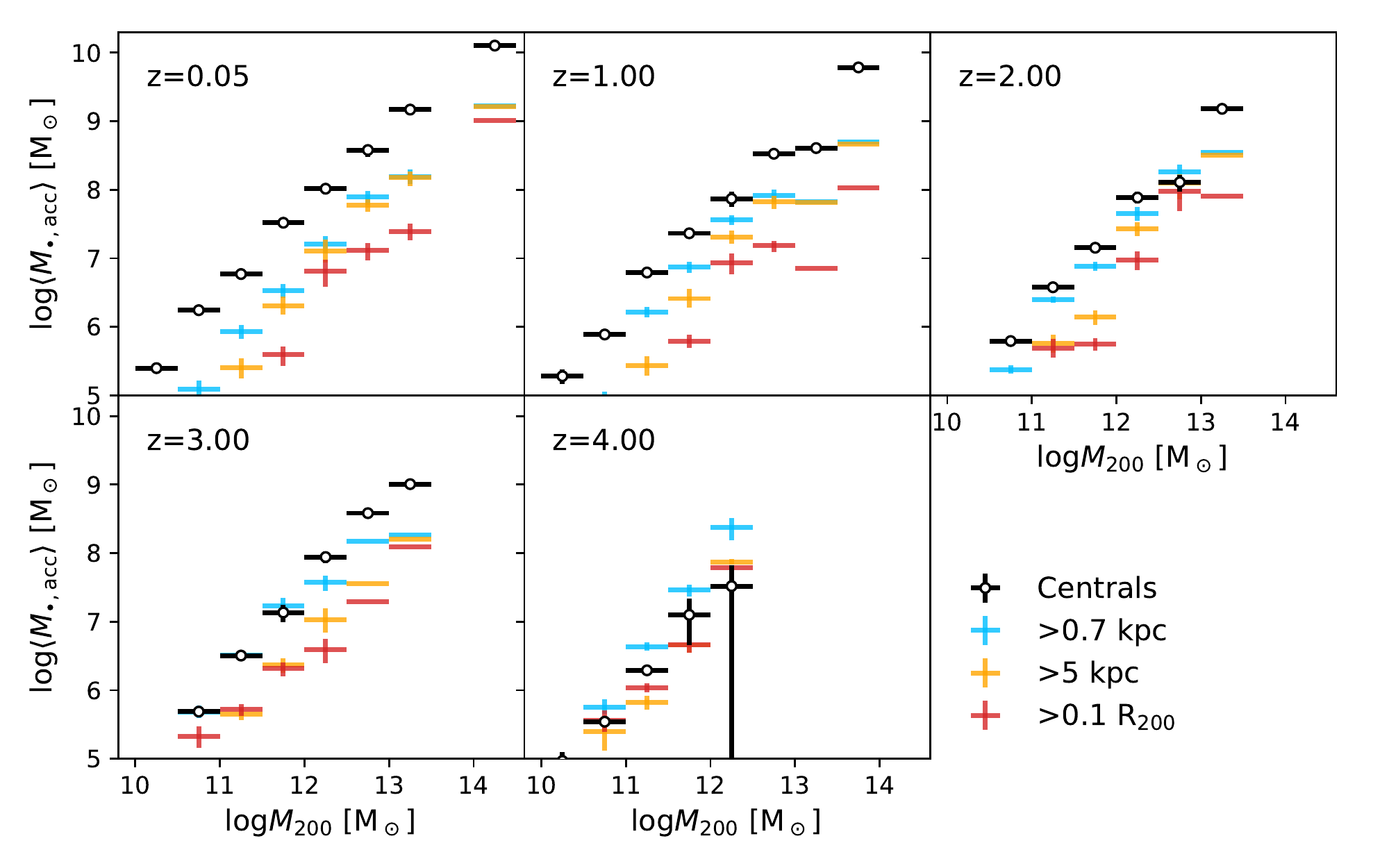}
  \caption{Average amount of mass in central and wandering SMBHs as a function of halo mass and redshift.  Only the accreted portion of SMBH masses are included, which excludes all seed masses.  The fraction of mass in wandering SMBHs is roughly 10 per cent the mass in central SMBHs at $z=0.05$, but this fraction increases with redshift.  At $z=4$, the mass locked in wanderers exceeds that locked in centrals, likely due to the increased merger rate at high redshift.  \label{fig:masses}}
\end{figure*}

To hone in on the growth history of these wanderers, we study the amount of mass locked up in wandering SMBHs as well as their accretion rates.  These estimates have important implications for the \citet{Soltan1982} argument, which connects the evolving AGN luminosity function (actively accreting SMBHs) to the total mass accumulated in relic central SMBHs as measured at $z=0$.  AGN emission is frequently used to estimate the cosmic SMBH growth rate at a given epoch, which when integrated from high redshift to the present day is set to equal to the mass of central SMBHs at $z=0$ \citep[e.g.,][]{Haehnelt+1998,Hopkins+2007,Merloni&Heinz2008,Shen+2020,Ananna+2020}.  Mass locked in wandering SMBHs is not accounted for in the Soltan argument and therefore represents a missing term from the right hand side of the equation.

In Figure \ref{fig:masses}, we first plot the average total mass included in {\it all} central or wandering SMBHs as a function of halo mass in {\sc Romulus25}.  Only the accreted portion of a SMBH's mass is included, removing the masses from all seeds which have merged to form the final product, which is the direct comparison for the Soltan argument.  As shown in \citet{Ricarte+2019}, the accreted mass is strongly correlated with the host stellar mass even when the accreted mass falls below the initial seed mass of $10^6 \ \mathrm{M}_\odot$.  Since most wandering SMBHs do not grow substantially from their initial seed mass, this also allows us to better differentiate the population of wanderers in terms of mass and allows us to make conservative estimates of the amount of mass locked up in wanderers.  

As with their total numbers, both central and wandering SMBH masses follow linear trends with halo mass.  At $z=0$, note that the sub-grid parameters of these simulations are tuned to reproduce the empirically measured central SMBH mass to stellar mass relation \citep[specifically][]{Schramm&Silverman2013}, and hence a linear relationship between $M_{\bullet,\mathrm{acc}}$ and $M_{200}$ for central SMBHs exists by construction.  Nevertheless, it is interesting that the linear relationship extends down to halos of even $10^{10} \ \mathrm{M}_\odot$, far below the mass threshold above which AGN feedback is thought to quench galaxies and shape their properties.  Overall, we find that wandering SMBHs produce only a roughly 10 per cent correction to the total amount of accreted mass locked up in SMBHs at $z=0$, which is small compared to the systematic uncertainties on scaling relations used to estimate the total mass locked up in SMBHs \citep{Kormendy&Ho2013,Reines&Volonteri2015,Saglia+2016,Shankar+2016}.

However, we find that as redshift increases, the fraction of mass tied up in SMBH wanderers increases, such that wandering SMBHs actually dominate the black hole mass budget at $z=4$.  This is despite the fact that dynamical times are shorter at high redshift \citep[e.g.,][]{Barkana&Loeb2001}.  This suggests that the increased merger rate at high redshift is reflected in and results in the existence of wandering SMBHs.  Higher redshift galaxies simulated in a cosmological environment are also clumpier, which can disrupt the angular momentum loss of wanderers via scattering events \citep{Pfister+2019,Bortolas+2020,Ma+2021}.

\subsubsection{Central and Wanderer Luminosities}

\begin{figure*}
  \centering
  \includegraphics[width=\textwidth]{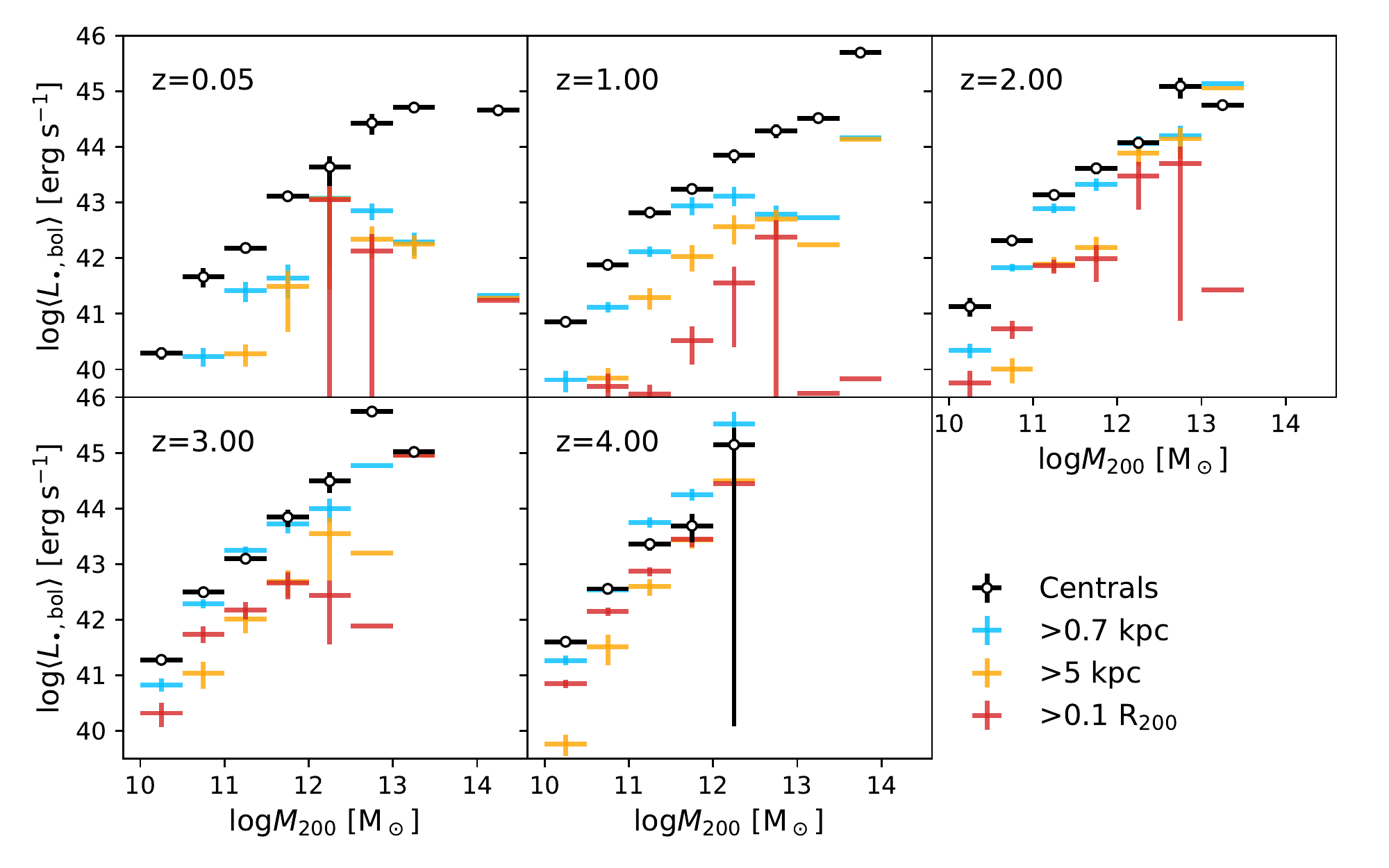}
  \caption{Average luminosities of central and wandering SMBH populations as a function of halo mass.  Wandering SMBHs outshine centrals in massive halos at $z=4$.  Wandering luminosities are suppressed in massive halos at low redshift, most likely due to AGN feedback from the central SMBH.  \label{fig:luminosities}}
\end{figure*}

We now examine if this wandering population implies and results in a corresponding increase in detectable emission.  A variety of models predict that wandering SMBHs can manifest as off-nuclear X-ray or radio sources \citep{Fujita2008,Fujita2009,Bellovary+2010,Sijacki+2011,Kawaguchi+2014,Steinborn+2016,Barrows+2019,Zivancev+2020,Guo+2020,Bartlett+2021}.  In Figure \ref{fig:luminosities}, we plot the average total bolometric luminosities of {\it all} central or wandering SMBHs as a function of halo mass in {\sc Romulus25}.  In order to mitigate AGN variability, luminosities are averaged over 30 Myr.  Self-consistently with the feedback prescription used in {\sc Romulus}, we assume a universal radiative efficiency of 10 per cent for all sources; $L_\mathrm{bol} = 0.1 \dot{M}_\bullet c^2$, where $\dot{M}_\bullet$ is the SMBH growth rate and $c$ is the speed of light.

Similar to the mass budget, we find that wandering SMBHs beyond 0.7 kpc can indeed contribute a significant amount to the total AGN photon budget of a halo, growing with redshift.  As suggested by their mass budgets, wandering SMBHs cumulatively outshine central SMBHs at $z=4.0$.  However, we also notice an interesting change in behaviour in the most massive halos at low redshift.  Although both the total number and total mass of wandering SMBHs continues to increase with halo mass, their cumulative accretion rates begin to decline in the group and cluster halos.  This is likely a consequence of AGN feedback from the central SMBH heating the gas in the most massive halos.  Unfortunately, this means that despite their high abundance, wandering SMBHs will be more difficult to detect via emission in group and cluster environments.

\subsection{Intra-halo Statistics}
\label{sec:intra-halo}

\subsubsection{The (Accreted) Mass Function of Wandering SMBHs}

\begin{figure*}
  \centering
  \includegraphics[width=\textwidth]{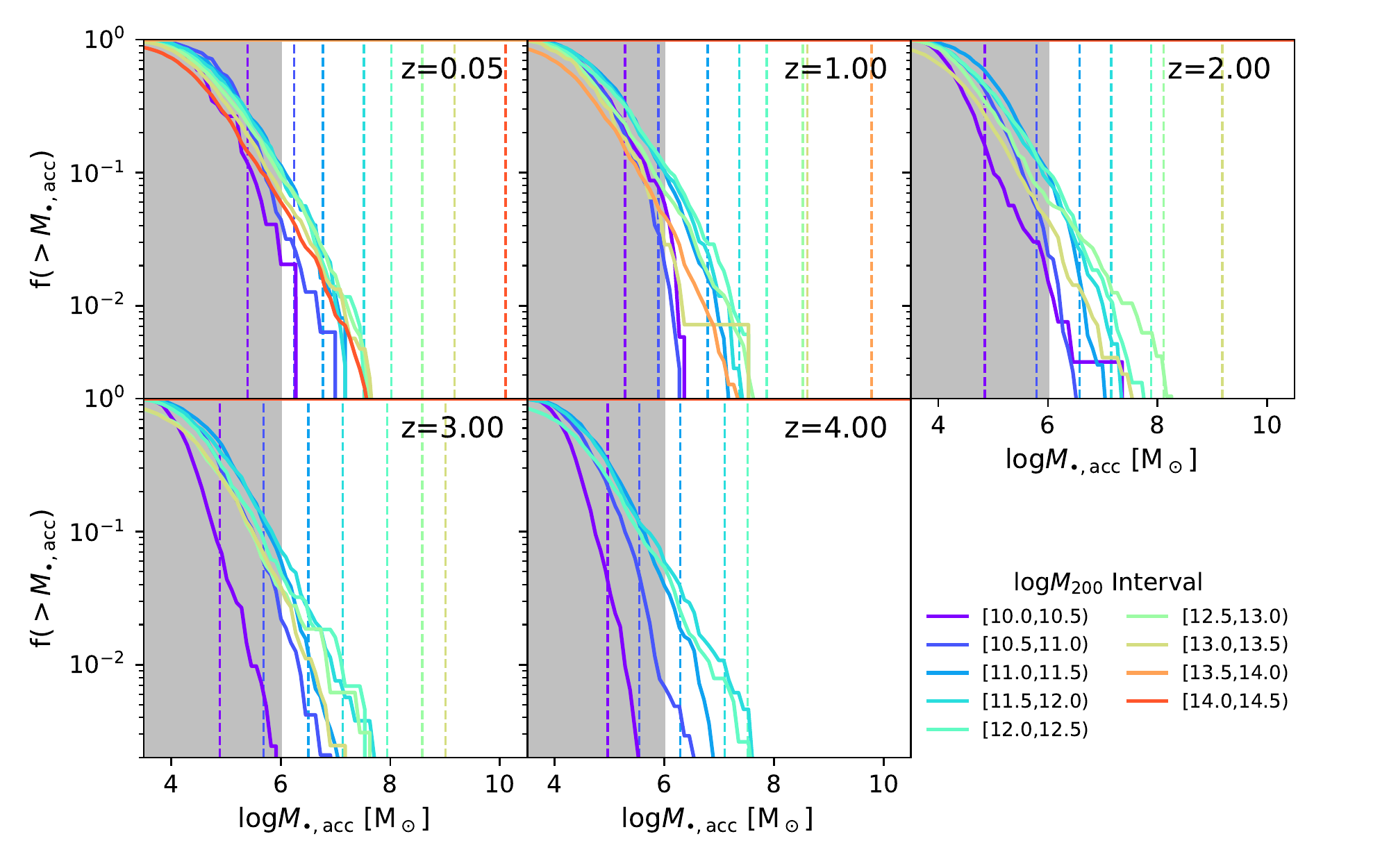}
  \caption{Mass function of wandering SMBHs as a function of halo mass and redshift.  Only the accreted portions of mass are included, which excludes seed masses.  Vertical dashed lines plot the median {\it central} SMBH mass at a given halo mass.  Mass functions of wandering SMBHs are very similar as a function of halo mass and redshift, even though central SMBH masses are highly halo mass dependent.  The vast majority of wandering SMBHs have accreted masses below the seed mass of $10^6 \ \mathrm{M}_\odot$ (within the grey coloured region). \label{fig:massFunction}}
\end{figure*}

Here, we explore the mass functions of wandering SMBHs in Figure \ref{fig:massFunction}.  \citet{Banik+2019} found a Schechter-like wandering SMBH mass function in {\sc Romulus25}, and here we search for trends in halo mass and redshift.  We again plot only the accreted portion of a SMBH's mass, excluding the initial seed mass.  The y-axis represents the fraction of SMBHs with accreted masses above that listed on the x-axis.  Different colours represent different halo masses, and the dashed vertical lines plot the median central SMBH mass in a given halo mass bin.  We find that the mass function of wandering SMBHs is dominated by SMBHs with accreted masses less than the seed mass of $10^6 \ M_\odot$.  That is, most of these wandering SMBHs never grew by a substantial fraction of their initial masses in the simulation.  Less than approximately 10 per cent of them have accreted masses greater than the seed mass.  

The universality of these curves is remarkable.  Despite the median central mass increasing dramatically with halo mass, the mass function of wandering SMBHs remains nearly the same.  This behaviour can be anticipated based on Figures \ref{fig:occupation} and \ref{fig:masses}:  both the number of wanderers and their cumulative masses are roughly linear with halo mass, implying that the typical wanderer mass does not change.  We do note, however, that more massive halos exhibit a larger tail of massive wanderers.  This is because, as we will further explore, wandering SMBHs originate from minor mergers.  

\subsubsection{Spatial Distributions}
\label{sec:radial_distribution}

\begin{figure*}
  \centering
  \includegraphics[width=\textwidth]{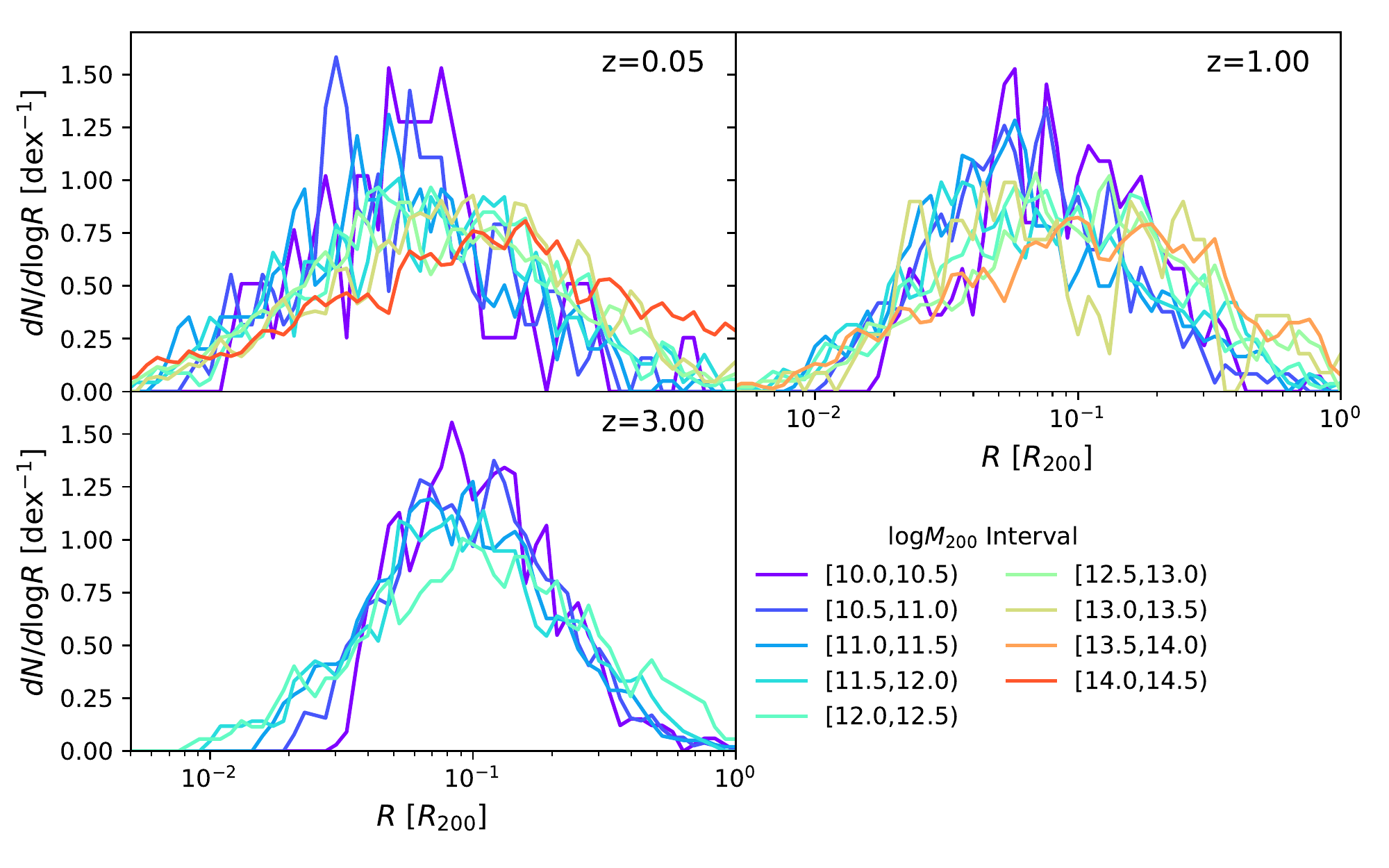}
  \caption{Spatial distributions of wandering SMBHs in the {\sc Romulus} simulations as a function of virial radius.  Different colours correspond to different bins in halo mass.  We find a broad distribution of radii with a peak at roughly a tenth the virial radius that moves outward as halo mass increases.  Different panels depict different redshifts, over which we do not see substantial evolution. \label{fig:distances}}
\end{figure*}

In Figure \ref{fig:distances}, we plot radial distributions of wandering SMBHs as a function of virial radius for three different redshifts.  The peaks of these distributions tend to occur at roughly at a tenth of the virial radius, moving outward with increasing halo mass.  There is a wide tail to large offsets that also increases with halo mass.  The spatial distribution of wandering SMBHs spans the full radial extent of the halo, although they remain much more centrally concentrated than the dark matter halo itself.  In the main {\sc RomulusC} halo at $z=0.05$ (the sole halo in the $10^{14-14.5} \ \mathrm{M}_\odot$ bin), a substantial population of wanderers can be found all the way out to the virial radius.  This distribution is robust and does not appear to change substantially with redshift.

\begin{figure*}
  \centering
  \includegraphics[width=\textwidth]{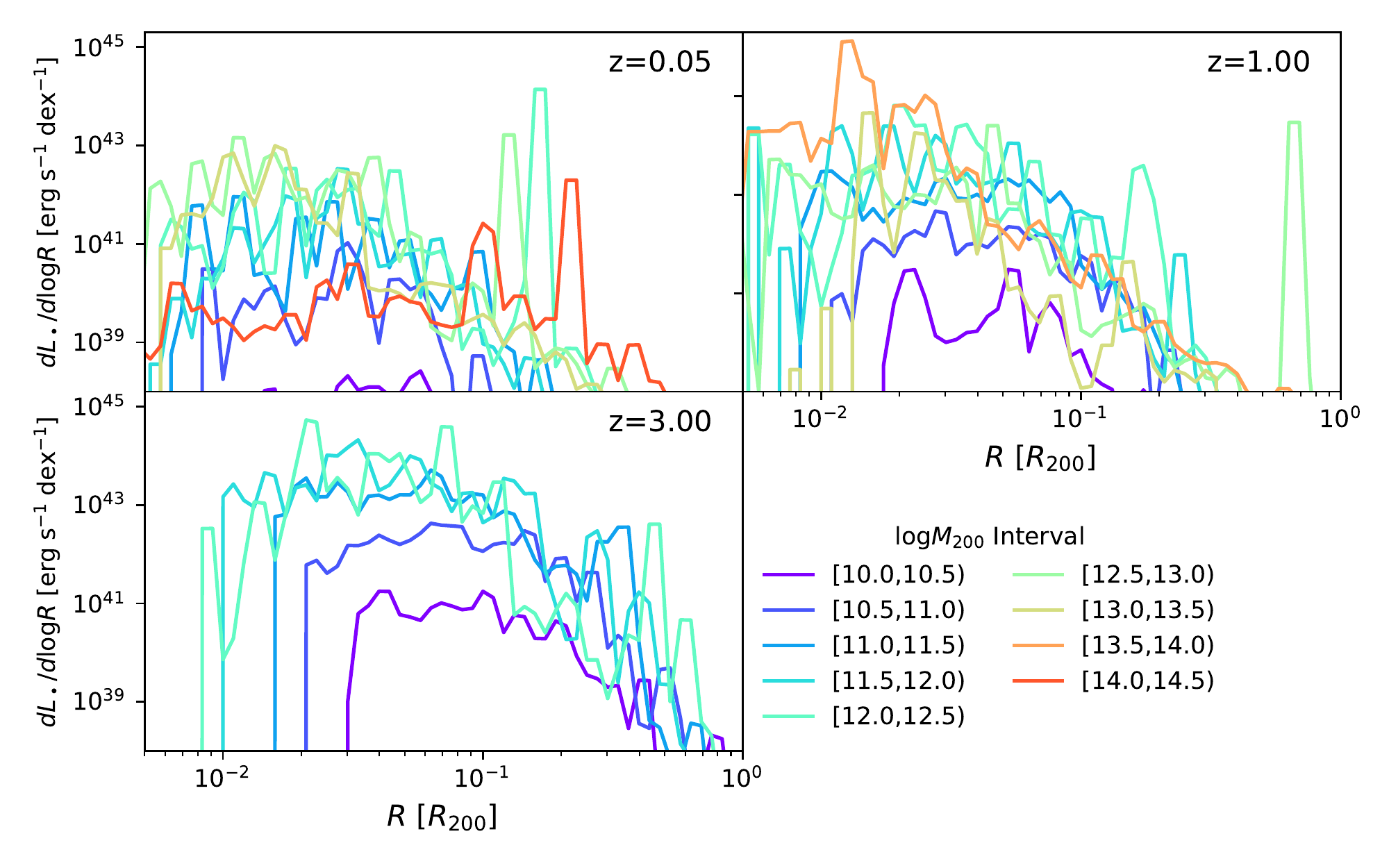}
  \caption{Bolometric luminosity distributions of wandering SMBHs as a function of virial radius for three different redshifts, assuming a radiative efficiency of 0.1.  Comparing with Figure \ref{fig:distances}, these distributions are more centrally concentrated than the population overall, since gas is both denser and colder at smaller radius. \label{fig:luminosity_profile}}
\end{figure*}

Unsurprisingly, the emission generated by these wandering SMBHs is more biased towards smaller radius than the population overall.  At smaller radius, the ambient gas is both denser and colder.  This is shown in Figure \ref{fig:luminosity_profile}, where we plot the average luminosity profile of halos of different mass for comparison with Figure \ref{fig:distances}.  Notice again the depression of the luminosity in the most massive bin compared to much less massive halos.

\begin{figure*}
  \centering
  \includegraphics[width=\textwidth]{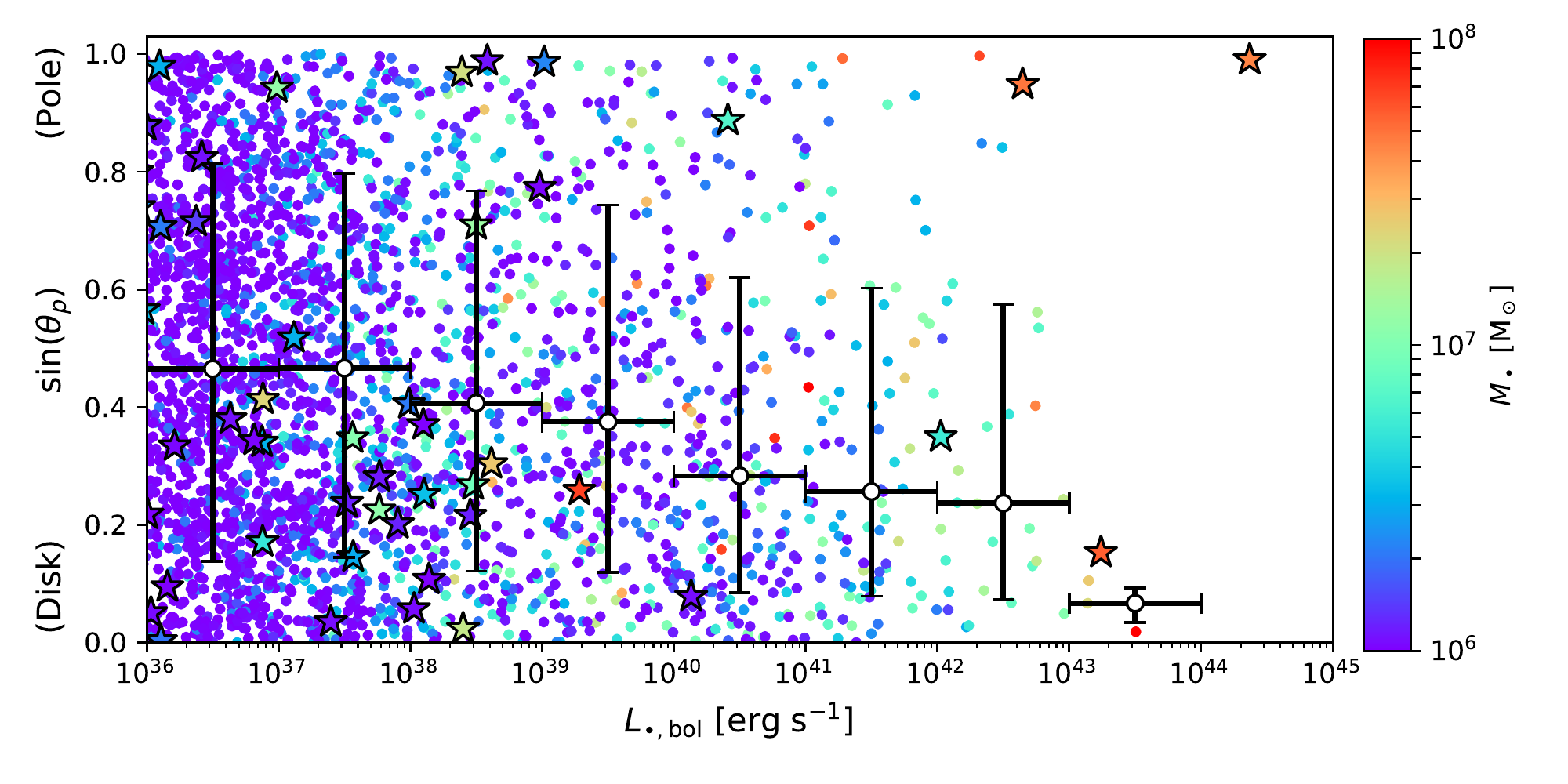}
  \caption{For each wandering SMBH at $z=0.05$, we plot $\sin(\theta_p)$ as a function of its bolometric luminosity, where $\theta_p$ is the angle made between the SMBH's position vector and the plane of the disk.  Points are coloured according to each SMBH's mass. Points marked by stars instead of circles correspond to objects residing in stellar over-densities of a factor of 10 or greater.  Error-bars span the 16th, 50th, and 84th percentile regions of the objects that do not reside in such over-dense regions.  This analysis reveals that the more luminous SMBHs are more likely to be more massive and closer to the plane.  \label{fig:wandererAngles}}
\end{figure*}

A deeper analysis reveals additional factors that promote growth onto wandering SMBHs.  In Figure \ref{fig:wandererAngles} we plot $\sin(\theta_p)$ as a function of bolometric luminosity (averaged over 30 Myr) for each wandering SMBH at $z=0.05$.  Here, $\theta_p$ is the angle made between each SMBH's position vector (from the center of its host halo) and the disk.  (By plotting $\sin(\theta_p)$ instead of $\theta_p$, we ensure that the y-axis evenly samples solid angle.)  If we detect a stellar overdensity of at least a factor 10 (see \S\ref{sec:stellar_counterparts} for details), we mark these SMBHs with stars instead of circles.  The errorbars represent the 16th, 50th, and 84th percentile regions, revealing that the more luminous wandering SMBHs are more likely to be close to the disk plane, as expected from previous studies \citep{Volonteri&Perna2005,Fujita2009}.  In addition, colours encode SMBH mass, revealing that the more luminous wanderers are also more likely to be more massive, as expected from the Bondi accretion formula.  When passing through the galactic disk, wandering SMBHs may also produce an observable bow shock \citep{Wang&Loeb2014}.  

\subsection{Do Wanderers Retain Stellar Counterparts?}
\label{sec:stellar_counterparts}

We explore whether or not wandering SMBHs reside in stellar over-densities by examining the particle data at redshift $z=0.05$.  If wandering SMBHs retain a stellar counterpart, this can make them easier to find.  This also serves as a test of our halo finder, which may have difficulty identifying low-mass halos in crowded environments.

We define the local stellar mass density $\rho_{*,\mathrm{local}}$ to be the mass density of stars within a 1 kpc radius sphere around a wandering SMBH.  Then, for comparison, we compute the typical stellar mass density of the SMBH's host halo at comparable radius and disk height, $\rho_{*,\mathrm{halo}}$.  To be more explicit, we select star particles within the intersection of (i) a spherical shell that encloses but excludes the aforementioned 1 kpc radius sphere, and (ii) the region defined by $|z-z_\bullet| \leq 1 \ \mathrm{kpc}$, where $z_\bullet$ is the vertical distance that the wandering SMBH is away from the galactic mid-plane, located using {\sc Pynbody}'s ``faceon'' function \citep{Pontzen+2013}.  We exclude wandering SMBHs that reside in halos for which this function fails to determine an angular momentum vector, which amounts to 20 per cent of the total number of wanderers.

\begin{figure*}
  \centering
  \includegraphics[width=\textwidth]{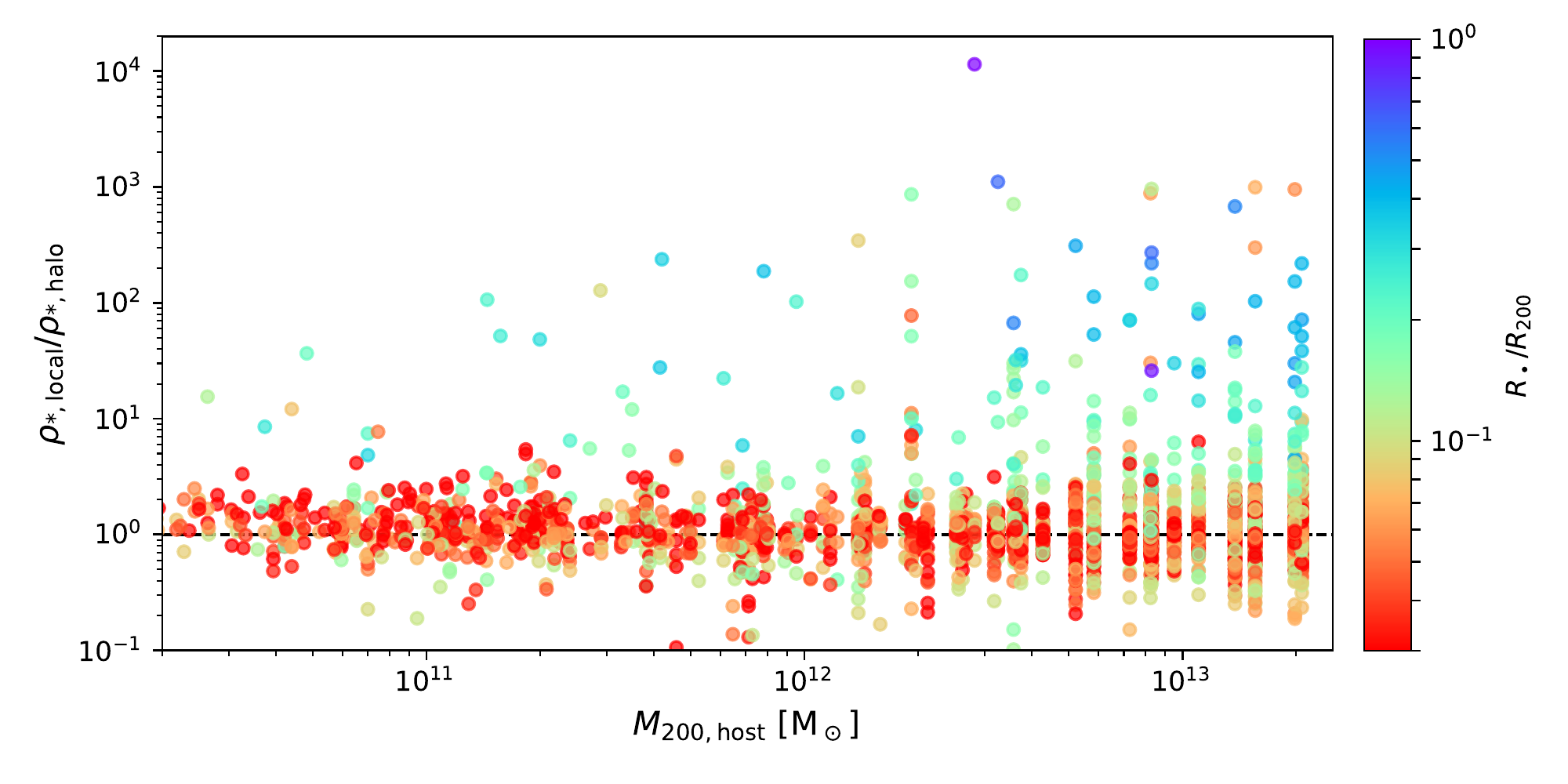}
  \caption{Local stellar over-density factors of wandering MBHs as a function of host halo mass at $z=0.05$.  Colours in this plot encode the ratio of each wanderer's halo-centric distance to the virial radius.  Most MBHs typically do not reside in stellar overdensities in {\sc Romulus25}, except for those at large halo-centric radius.  This is consistent with a picture in which wandering SMBHs originate from tidally disrupted satellite halos. \label{fig:stellarOverdensities}}
\end{figure*}

Results are plotted in Figure \ref{fig:stellarOverdensities}.  The ratio $\rho_{*,\mathrm{local}}/\rho_{*,\mathrm{halo}}$ is plotted as a function of host halo mass, $M_\mathrm{200,host}$.  Colours encode the ratio of the SMBH's halo-centric distance to the host virial radius.  Overall, the majority of wandering SMBHs appear to exist in ambient stellar densities typical for the halo overall, implying that the halos in which they formed have been completely tidal disrupted.  Some exceptions exist, however, and as revealed by the colouration, those SMBHs which do reside in stellar overdensities tend to exist a large halo-centric radius.  These SMBHs likely retain a stellar component due to the weaker tidal forces in the halo outskirts.  Fortunately, these cases are not likely to be due to halo finder inaccuracies, which would have greater difficulty identifying substructure at small radius rather than at large. 

The fact that wandering SMBHs typically do not exist in stellar overdensities except at large radius bolsters our confidence that we are not simply selecting for SMBHs in regions where the halo finder has a difficult time identifying substructure.  We note that the 350 pc resolution of this simulation disallows the formation of potentially denser stellar structures like nuclear star clusters, which are found in approximately 70 per cent of galaxies with less than $10^{11} \ \mathrm{M}_\odot$ in stars \citep{Neumayer+2020}. Nuclear star clusters may well be the birthplaces of SMBH seeds \citep{Devecchi&Volonteri2009,Alexander&Natarajan2014,Natarajan2021}, and may also be an additional source of wanderers if they form offset from their galactic centres.  Such dense stellar structures can more easily avoid tidal disruption and also help SMBHs to sink more efficiently to the centre of galaxies \citep{vanWassenhove+2014,Tremmel+2018a}.  Finally, numerical disruption is known to artificially destroy the innermost bound clumps of halos experiencing tidal stripping, which may also cause {\sc Romulus} to underestimate the local stellar and dark matter density around wandering SMBHs \citep{vandenBosch+2018}.

\section{Discussion}
\label{sec:discussion}

From our detailed study, we report that wandering SMBHs are plentiful in the {\sc Romulus} universe, with numbers scaling linearly with the halo mass.  Their masses imply only a roughly 10 per cent correction to the total mass locked up in SMBHs locally, but they may collectively outweigh the contribution from central SMBHs at high redshifts when galaxy mergers are more frequent.  We find that wanderers typically have very low masses (similar to their initial seed masses) and do not typically reside in halo over-densities, with a large fraction lying beyond the visible extent of their hosts.  Here, we discuss some of the numerical limitations and broader implications of this study.

\subsection{The SMBH Seeding Mechanism}

Due to limited resolution, cosmological simulations cannot seed SMBHs at masses lower than those of the other particles in the simulation.  The {\sc Romulus} seeding prescription is based on the direct collapse black hole mechanism, which may produce ``heavy'' seeds up to $\approx 10^6 \ M_\odot$ \citep[e.g.,][]{Begelman+2006,Lodato&Natarajan2007}, which is the seed mass employed in these simulations.  Alternative ``light'' seeding mechanisms also likely operate in the Universe in tandem with or instead of a ``heavy'' seeding mechanism \citep[e.g.,][]{Madau&Rees2001,Volonteri+2003}.  Light seeds, such as the remnants of the first Pop III stars, could be established in greater numbers, but with lower initial masses.  Under a light seeding scenario, we would expect an even greater number of wandering SMBHs, with a mass function extending to lower masses than the $10^6 \ \mathrm{M}_\odot$ seeds found in {\sc Romulus}, as explored in detail in \citep{Rashkov&Madau2014}.  We reiterate, however, that the heavy seeding mechanism employed by {\sc Romulus} successfully reproduces the observed local SMBH occupation fraction, as shown in \S\ref{sec:numbers}. Thus, in terms of the census of SMBHs both central and wandering, we believe that the {\sc Romulus} simulations offer a plausible picture, even though we are likely missing BHs that originated and grew from light seeds.

\begin{figure}
  \centering
  \includegraphics[width=0.5\textwidth]{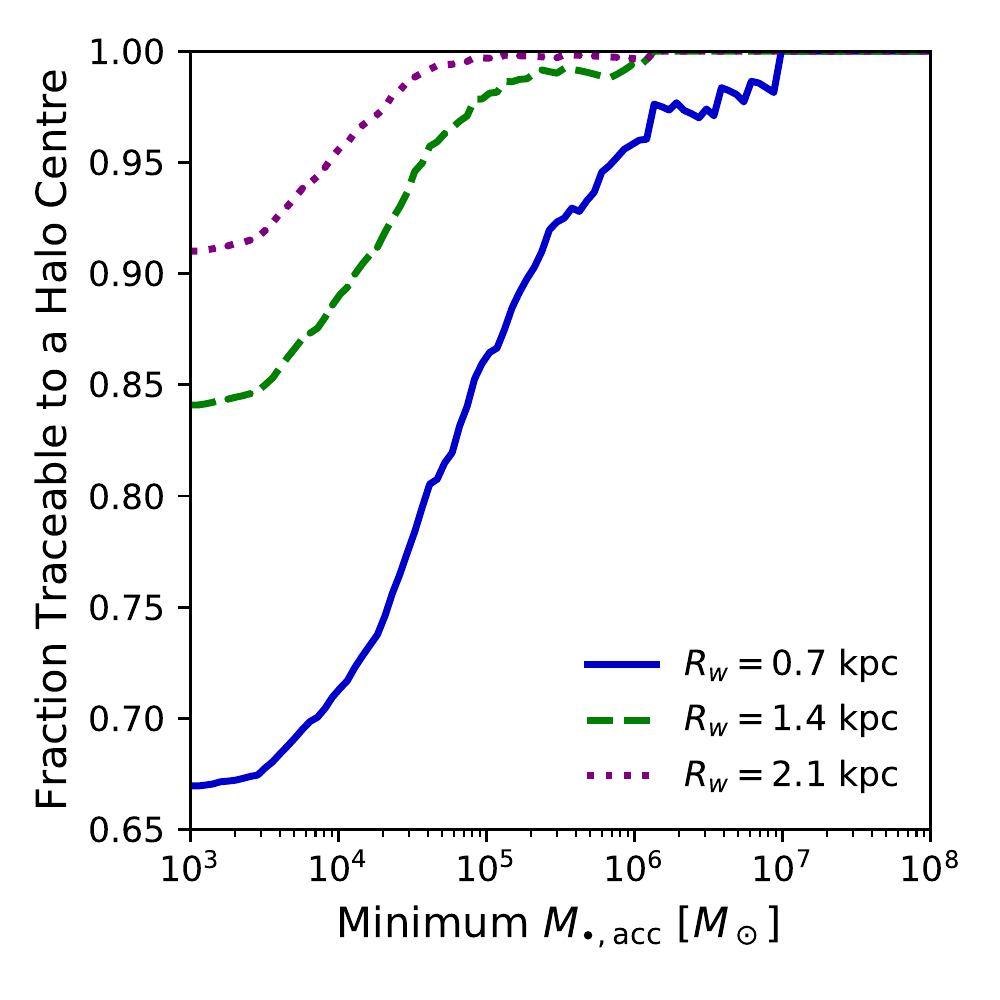}
  \caption{Fraction of wandering SMBHs at $z=0.05$, above the accreted mass indicated on the x-axis, that can be traced back to the centre of any halo.  Three values of the halo-centric radius defining the centre of the halo, $R_w$, are shown.  The vast majority of SMBHs can be traced to the centres of halos at some point earlier in their histories. This demonstrates that {\sc Romulus} does not create spurious seeds, and our wandering SMBHs were once located at the centres of destroyed halos. \label{fig:historicalDistanceCheck}}
\end{figure}

Since the {\sc Romulus} seeding prescription is based on local gas properties, one might worry that the wandering SMBHs we find in the simulation may come from spurious seeding sites.  To test this, and to gain a better understanding of their origins, we trace back in time the wandering SMBHs at $z=0.05$ and examine their host halos at previous time steps.  In Figure \ref{fig:historicalDistanceCheck}, we plot the fraction of SMBHs, above a minimum accreted mass plotted on the x-axis, which can be traced to a halo centre at any point in their histories.  To be more explicit, for a halo-centric radius $R_w$ beyond which a SMBH is considered wandering, we plot the fraction of wanderers at $z=0.05$ that were {\it not} wandering at any point in their histories.  We confirm that the vast majority of wandering SMBHs originate from the centres of resolved halos at earlier time slices.  This fraction decreases with decreasing accreted mass, meaning that wanderers with greater masses were more likely to have been closer to the centres of a host halo in the past.  For the wandering population overall, we report that for $R_w$ of $0.7$ kpc, $1.4$ kpc, and $2.1$ kpc, the fractions traceable to the halo centre are 67 per cent, 84 per cent, and 91 per cent.  In other words, less than 10 per cent of wandering SMBHs have {\it never} been within 2.1 kpc of a halo centre.  It is even possible for some of wanderers to have been seeded in halos that formed and were destroyed and/or perturbed by a merger or flyby event in the time between snapshots, causing us to underestimate the fraction traceable to halo centres. Similarly, it is possible for seeds to form in the centres of overdensities too small and with too few particles to be considered resolved by our strict halo finding criteria.  Again, despite the fact that the {\sc Romulus} seeding prescription does not use global halo or galaxy properties, it naturally places SMBHs in halo centres.

\subsection{Additional Mechanisms to Produce Wanderers}

Due to limited resolution, two additional mechanisms that could potentially produce wanderers are neglected in the Romulus simulations:  multi-body SMBH interactions \citep{Volonteri+2003,Hoffman&Loeb2007,Ryu+2018}, and gravitational wave recoil from BH mergers \citep{Libeskind+2006,Volonteri2007,Lousto+2011,Blecha+2016}.  We argue that these additional channels are unlikely to significantly change the average number of wandering SMBHs at a given halo mass.  This is because these processes require major galaxy mergers to first form a bound SMBH binary, whereas the wanderers studied in this work originate mainly from prior minor mergers.  Since minor mergers always outnumber major ones, the {\sc Romulus} simulations should capture the majority of wandering SMBH population.  However, semi-analytic models predict that ejection events can dominate the wandering population at low halo-centric radius \citep{Volonteri&Perna2005,Izquierdo-Villalba+2020}.

\subsection{Bondi Accretion}

Perhaps the greatest uncertainty in the observability of wandering SMBHs is our assumption of Bondi accretion.  Recent 3D hydrodynamical simulations suggest that wandering SMBHs fed with hot and diffuse plasma with density fluctuations are limited to 10-20 per cent of the Bondi-Hoyle-Lyttleton accretion rate due to a wide distribution of inflowing angular momentum \citep{Guo+2020}.  In addition, the accretion rate can be suppressed by several additional orders of magnitude below the Bondi rate due to the accumulation of magnetic flux near the horizon in advection dominated accretion flows (ADAFs) \citep{Igumenshchev&Narayan2002,Perna+2003,Pellegrini2005,Ressler+2021}. 

\begin{figure*}
  \centering
  \includegraphics[width=\textwidth]{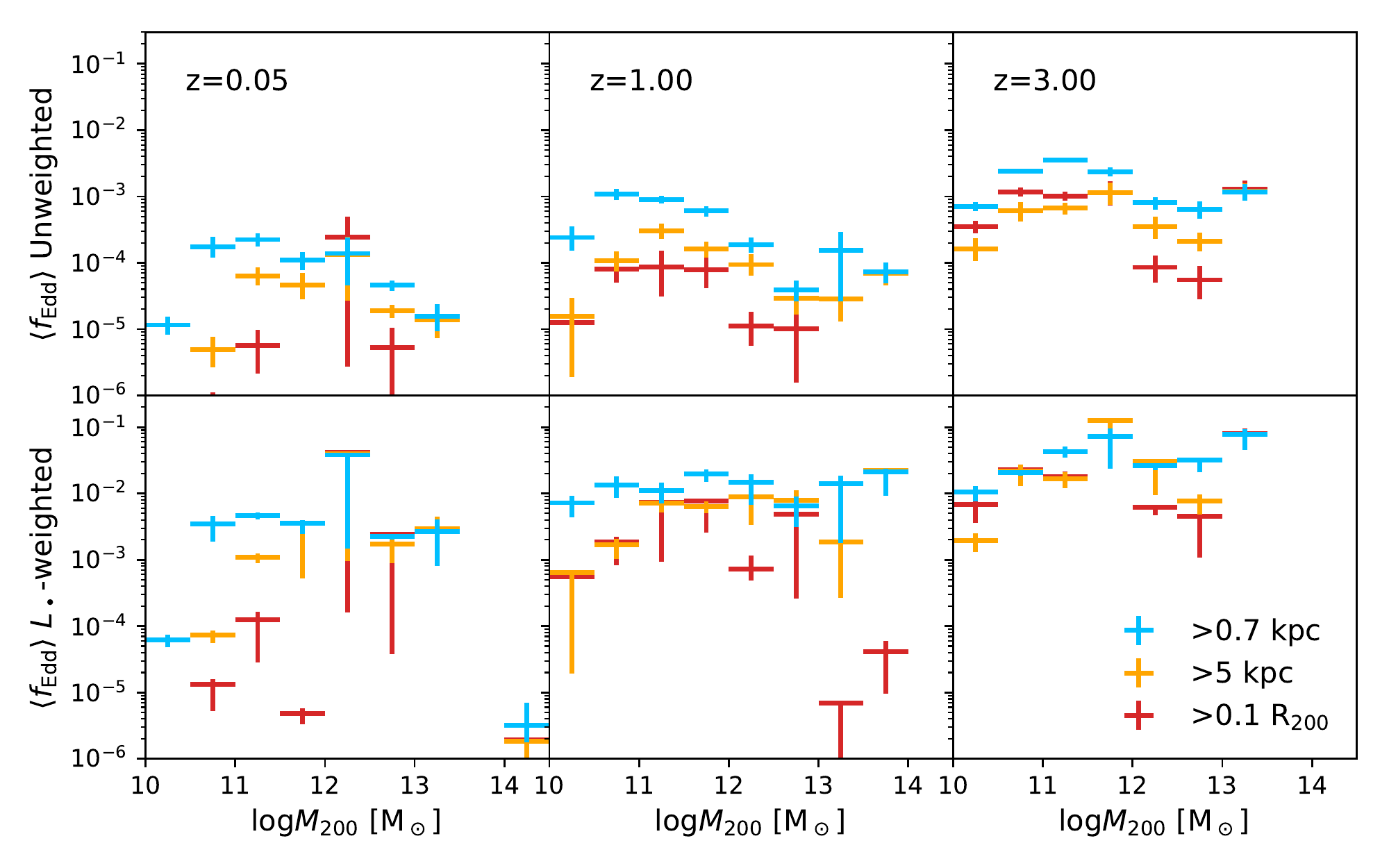}
  \caption{Average Eddington ratios of wandering SMBHs as a function of halo mass for three different redshifts. In the top row, each wandering SMBH is treated equally, while in the bottom row, we weight each wandering SMBH by its accretion rate.  The typical wandering SMBH has a modest Eddington ratio of $10^{-4}$, but the ones producing detectable emission have Eddington ratios orders of magnitude larger.  The typical Eddington ratio increases with redshift, as the universe becomes denser and more gas-rich.  \label{fig:eddington}}
\end{figure*}

To assess the potential severity of this second effect, we compute the typical Eddington ratios of wandering SMBHs.  ADAFs are believed to form around SMBHs accreting at Eddington ratios of $f_\mathrm{Edd} \lesssim 10^{-3}$, at which point the disk becomes geometrically thick with a lower radiative efficiency \citep{Ichimaru1977,Rees+1982,Narayan&Yi1994,Narayan&Yi1995,Abramowicz+1995,Reynolds+1996a,Yuan&Narayan2014}.  The Eddington ratio is given by $f_\mathrm{Edd} = \dot{M}_\bullet/\dot{M}_\mathrm{Edd}$ for $\dot{M}_\mathrm{Edd} = 4\upi G M_\bullet m_p / \epsilon \sigma_T c$, where $G$ is the gravitational constant, $m_p$ is the proton mass, $\sigma_T$ is the Thomson cross-section, $c$ is the speed of light, and $\epsilon$ is the radiative efficiency.  $\dot{M}_\mathrm{Edd}$ is the mass accretion rate above which radiation pressure inhibits the accumulation of greater mass flux.

We plot the average Eddington ratios as a function of halo mass in Figure \ref{fig:eddington} for our different cuts of the wandering population.  The top row treats every wandering SMBH equally, whereas the bottom row weights each SMBH according to its accretion rate (averaged over 30 Myr), which is more relevant when considering the emitting population.  We find that the typical wandering SMBH exhibits a modest Eddington ratio of $\sim 10^{-4}$ locally, but that this value increases with increasing redshift, when the universe is denser.  However, the emitting population is characterised by much higher Eddington ratios, around $10^{-2}$ at high redshift.  These values may be high enough to avoid the transition to ADAF and suppression of the Bondi rate due to magnetic fields.  We also note that the {\sc Romulus} simulations most likely under-predict the frequency of high Eddington ratio systems, instead producing slow and steady growth proportional to the star formation rate \citep{Ricarte+2019}.

\section{Conclusions}
\label{sec:conclusions}

We have performed a census of wandering SMBHs in the {\sc Romulus} cosmological simulations.  These simulations carefully correct the dynamical friction force onto SMBHs and produces occupation fractions consistent with observations using only local seeding conditions.  Our results are summarised below:

\begin{itemize}
    \item Wandering SMBHs in the {\sc Romulus} simulations originate from the centres of destroyed infalling satellite galaxies.  By $z=0.05$, they do not typically retain a resolvable stellar component in these simulations, but those which do exist at large fractions of their host halo's virial radius.
    \item These simulations predict on the order of 10 wandering SMBHs in Milky Way-mass halos at $z=0$, and their number scales linearly with halo mass, up to 1613 wandering SMBHs in the $10^{14} \ \mathrm{M}_\odot$ main halo of the {\sc RomulusC} cluster simulation.
    \item The total mass in wandering SMBHs is roughly 10 per cent the total mass in central SMBHs at $z=0.05$, but this fraction grows with redshift and even exceeds unity for $z>3$.  Similarly, the total luminosity budget of accreting SMBHs at $z>3$ is dominated by wanderers.  This is likely due to the increased merger rate at high redshift.
    \item Wandering SMBHs remain at large radius throughout the dark matter halos of their hosts.  Only roughly half of wandering SMBHs are within 10 per cent of the virial radius.  However, those with larger accretion rates are more likely to be at lower radius, higher mass, and closer to the galactic plane.
    \item Wandering SMBHs in {\sc Romulus} originate from the centres of halos destroyed in minor mergers.  The population is dominated by SMBHs at the seed mass regardless of host halo mass.  Most do not retain a resolved stellar counterpart, but those that do tend to be at large fractions of the virial radius, where tidal effects are weaker.
\end{itemize}

The {\sc Romulus} simulations predict a substantial wandering SMBH population, which may even dominate AGN emission at large redshifts.  In future work, we will explore the observational signatures of the wandering population in greater detail.

\section{Acknowledgements}

AR thanks Ramesh Narayan for continued support and insightful conversations about wandering SMBH accretion rates.  AR is supported by the National Science Foundation under Grant No. OISE 1743747 as well as the Black Hole Initiative (BHI) by grants from the Gordon and Betty Moore Foundation and the John Templeton Foundation. MT is supported by an NSF Astronomy and Astrophysics Postdoctoral Fellowship under award AST-2001810. The {\sc Romulus} simulations are part of the Blue Waters sustained-petascale computing project, which is supported by the National Science Foundation (awards OCI-0725070 and ACI-1238993) and the state of Illinois. Blue Waters is a joint effort of the University of Illinois at Urbana–Champaign and its National Center for Supercomputing Applications.  This work is also part of a Petascale Computing Resource Allocations allocation support by the National Science Foundation (award number OAC-1613674). This work also used the Extreme Science and Engineering Discovery Environment (XSEDE), which is supported by National Science Foundation grant number ACI-1548562. Resources supporting this work were also provided by the NASA High-End Computing (HEC) Program through the NASA Advanced Supercomputing (NAS) Division at Ames Research Center.  Analysis was conducted on the NASA Pleiades computer and facilities supported by the Yale Center for Research Computing.

\section{Data Availability}

The data used in the work presented in this article is available upon request to the corresponding author.

\bibliography{ms}

\end{document}